\def\etal{{et al.\thinspace}}
\def\sqdeg{deg$\rm^{2}$}
\def\Hi{\ifmmode \mathrm{H}\,\textsc{i}\else H\,\textsc{i}\fi}
\documentstyle[epsfig,psfig]{mn}

\begin{document}

\title[Reddening-Independent Quasar Selection]
      {Reddening-Independent Quasar Selection from a Wide-Field 
       Optical and Near-IR Imaging Survey}

\author[Sharp R.G. \etal]
       {R. G. Sharp,$^{1}$, C. N. Sabbey,$^{1,2}$, A. K. Vivas,$^{2}$
        A. Oemler, Jr,$^{3}$ R. G. McMahon,$^{1}$ \newauthor
        S. T. Hodgkin,$^{1}$ P. S. Coppi,$^{2}$\\ $^1$Institute of
        Astronomy, Madingley Road, Cambridge, CB3 0HA, UK\\
        $^2$Astronomy Department, Yale University, P.O. Box 208101,
        New Haven, CT 06520-8101 USA\\ $^3$Carnegie Observatories, 813
        Santa Barbara Street, Pasadena, CA 91101 \\
        e-mail: rgs@ast.cam.ac.uk, rgm@ast.cam.ac.uk\\ }

\date{Accepted ....  Received ....}

\maketitle

\begin{abstract}
We combine deep, wide-field near-IR and optical imaging to demonstrate
a reddening-independent quasar selection technique based on
identifying outliers in the (g $-$ z) / (z $-$ H) colour diagram.  In
three fields covering a total of $\approx 0.7$ deg$^2$ to a depth of
m$\rm_H$$\sim$18, we identified 68 quasar candidates.  Follow-up
spectroscopy for 32 objects from this candidate list confirmed 22
quasars (0.86$<$z$<$2.66), five with significant IR excesses.  2 of 8
quasars from a subsample with U band observations do not exhibit UVX
colours.  From these preliminary results, we suggest that this
combined optical and near-IR selection technique has a high selection
efficiency ($> 65$\% success rate), a high surface density of
candidates, and is relatively independent of reddening.  We discuss
the implications for star/galaxy separation for IR base surveys for
quasars.  We provide the coordinate list and follow-up spectroscopy
for the sample of 22 confirmed quasars.
\end{abstract}


\section{Introduction}
Quasar spectral energy distributions are diverse and comparing surveys
at different wavelengths, including the infrared, is important for
characterizing the quasar population.  In addition, compared to
optical surveys, infrared selection is less effected by dust
extinction and reddening (due to both dust within host galaxies or
along the line of sight).  This has implications not only for
characterizing the influence of dust obscuration on the observed
population, but also for understanding biases in lensing (Kochanek
1996) and Damped Lyman-$\alpha$ (DLA) absorption system studies based
on optical selected quasar samples (Pei and Fall 1995, Fall, Pei,
McMahon 1989).

Infra-red observations were first used in quasars selection by
Braccesi, Lynds and Sandage (1968) where they showed how Infrared
excess could be used in conjunction with UV excess, derived using
U$-$B colours, to distinguish quasars from galactic foreground
stars. These early observations used red sensitive photographic
emulsions which were sensitive to wavelengths beyond that of the human
eye but which we would now call the the optical I
band(7000-9000\AA). The underlying physical principle behind the
technique was the use of a wide wavelength range, 3500-8000\AA, large
enough to distinguish the black body dominated stellar spectra from
the non-thermal powerlaw dominated spectral energy distribution of
quasars.  The use of IR data alone to discover quasars was
demonstrated by Beichman \etal (1998) who discovered a z=0.147 quasar
using JHK data from the 2 Micron All Sky Survey (2MASS).  However, due
to current technological limitations, obtaining deep multi band IR
observations over a sufficiently wide field of view to allow the
construction of a sample of satisfactory size at cosmologically
interesting redshifts (i.e. z=1-2) is prohibitively expensive in terms
of telescope time. The availability of large infrared mosaic cameras
makes possible wide-field IR surveys and the selection of large quasar
samples at IR wavelengths.

The advantage of the IR approach compared to optical is the reduced
influence of both dust extinction and reddening.  The removal of
sources from a flux limited sample by dust extinction will be more
severe at optical wavelengths than in the infrared.  In addition, as
discussed below, the reddening vector does not push quasars into the
stellar locus.  For example, Francis, Whiting, and Webster 2000 show
that red quasars are indistinguishable from the stellar locus in
optical multi-colour surveys, but often can be distinguished with the
addition of near-IR data.  Similarly, Barkhouse and Hall 2001
recommend the combination of optical and near-IR as most effective
(verse optical or near-IR separately).

A complementary approach is to use radio or hard X-rays as a method to
identify AGN in a manner that is not strongly biased by dust. In a
recent program, Ellison \etal (2001) report the results of an survey
for DLAs within a sample of radio selected quasars.  Ellison
\etal (2001) used a sample of 878 flat spectrum radio source with
S$\rm_{2.7GHz}$$>$250mJy.  However it is not obvious that these
surveys can discover enough quasars with z$>$2, which are luminous
enough in the optical that searches for damped Lyman-$\alpha$ can be
carried out, since typically $\sim$1\% of quasars have radio emission
at the 250mJy flux level reached by Ellison \etal (2001).  Hooper
\etal (1995) observed 3 out of 256 quasars (1.2\%) with
m$\rm_B$$<$18.5 and S$\rm_{8GHz}$$>$250mJy , rising to 14 (5.5\%) with
S$\rm_{8GHz}$$>$ 25mJy). Using a deeper radio survey is possible but
at the mJy level, radio surveys are dominated by faint galaxies and
the follow-up of such surveys could be prohibitive.

Our approach is similar to the KX technique of Warren, Hewett, and
Foltz (2000) which proposes selecting quasar candidates by their
K-excess in a V$-$J/J$-$K colour-colour diagram (see Croom, Warren and
Glazebrook (2001) for an application of the method).  As an
alternative to VJK based candidate selection we demonstrate a
technique based on gzH observations.  There are two reasons for our
alternate band selection.

1) Only one near IR band is required.  The use of the z band, readily
   obtained with large format CCD based cameras, leads to an increase
   in the practicality of surveying a large area. The relatively small
   field of view of IR array cameras is imposed only on H band
   observations.

2) We demonstrate that after accounting for sky brightness,
   K-correction and potential differential dust extinction effects
   between the H and K bands, observations in either band are
   equivalent with respect to practical observational requirements.
   The current availability of wide area IR imaging observations in
   the H band with CIRSI (section \ref{Near-IR and Optical
   Observations}) make sample definition possible over a wide field of
   view (in excess of 10deg$^2$).

\begin{figure}
\psfig{file=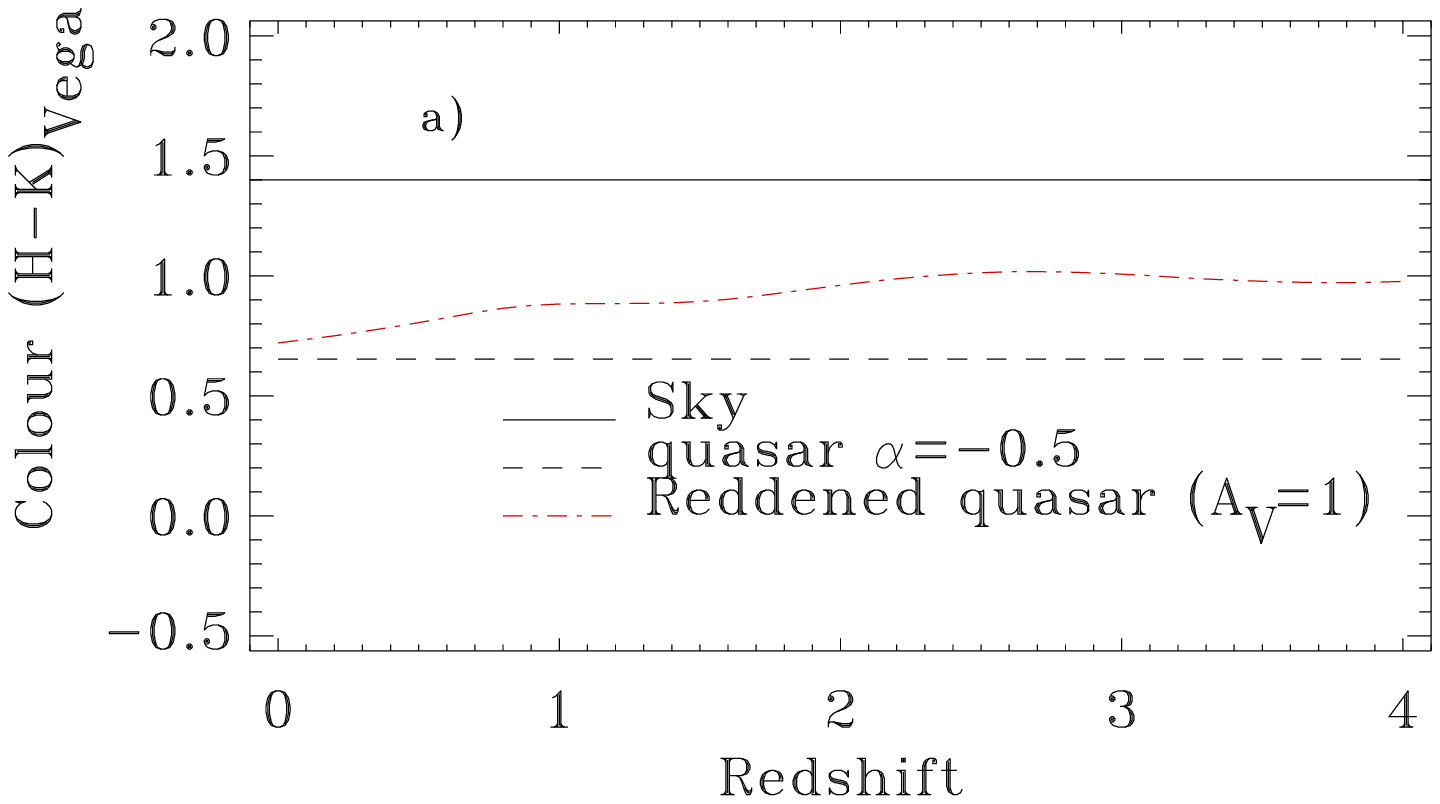,width=8cm}
\psfig{file=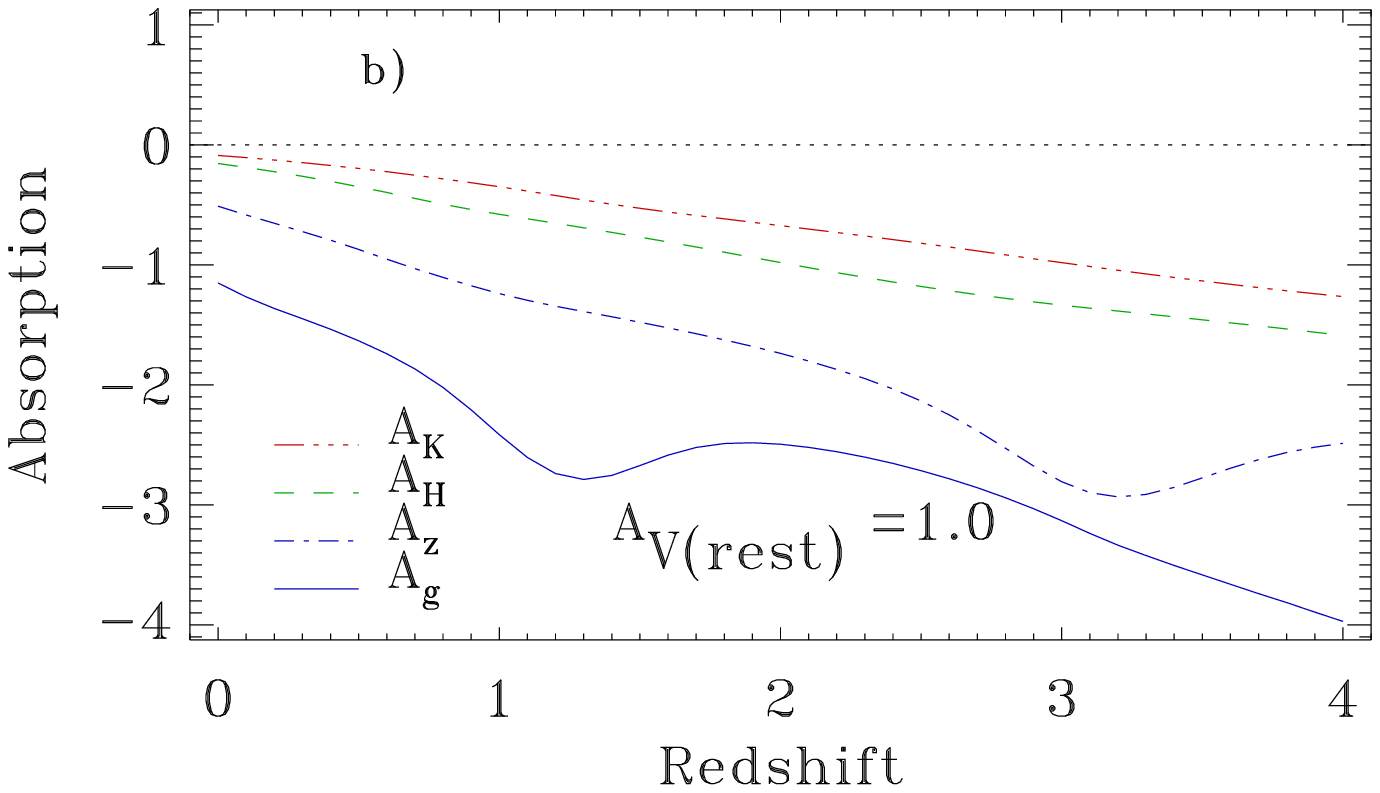,width=8cm}
\caption[]{\label{reddening} a) The (H$-$K) colour for a power law model quasar
spectrum ($\alpha$=$-$0.5) is compared to that of a power law with
reddening at the level of A$\rm_{V(rest)}$=1.0.  The (H$-$K) sky
colour is indicated by the solid line.\\ b) Absorption as a function
of redshift for A$\rm_V$=1.0 in the rest frame is shown in the gzH and
K bands.\\The galactic reddening law of Seaton 1979 is used.}
\end{figure}

Based on information obtained from the ESO ISSAC and SOFI exposure
time calculator (http://www.eso.org/observing/etc/) we find the H and
K sky brightnesses to be H=14.4 and K=13.0.  The quasar (H$-$K)
K-correction is estimate to be $-$0.6 (assuming a simple power law
quasar model $\alpha$=$-$0.5).  Under these assumptions quasars are
0.8mag brighter relative to the sky in H than in K.  Strong reddening
(A$\rm_{V(rest)}$$>$2.0) in the quasar rest frame or along the line of
site is required before this situation is reversed due to the shallow
slope of the reddening function at longer wavelengths under most
parameterization of the reddening law (figure \ref{reddening}).

Although several papers have used the colours of previously known
quasars to discuss quasar selection using combined optical and near-IR
imaging, we demonstrate this technique by carrying out a survey for
previously unknown quasars.  We present follow-up spectroscopic
observations that confirm the technique, and suggest high selection
efficiency, high surface density, and relative independence from
reddening.

In the following section (\ref{Near-IR and Optical Observations}) we
discuss the infrared and optical imaging data, section (\ref{colours})
describes the selection of quasar candidates using colour criteria,
section (\ref{host}) is concerned with the implications from detection
of the host galaxy.  The spectroscopic observations are described in
section (\ref{spectra}), and the sample of confirmed quasars is
presented in section (\ref{quasars}).

Unless stated otherwise we use conventional Vega magnitudes and
H$_{0}=$50 km s$^{-1}$ Mpc$^{-1}$, q$_{0}$=0.5 throughout this work.

\begin{figure}
\psfig{file=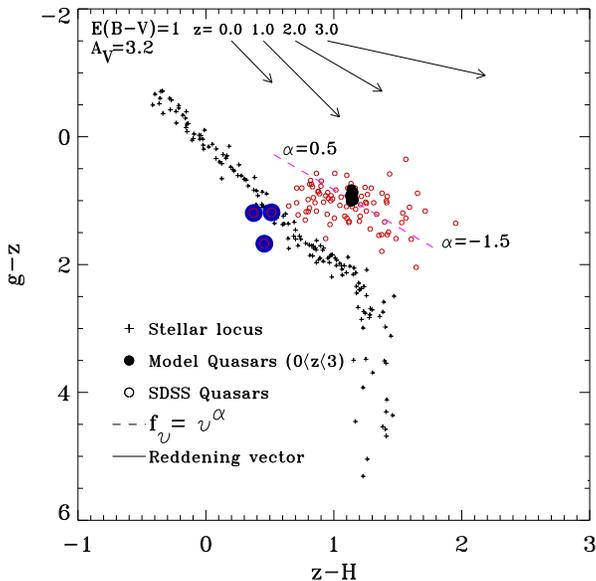,width=8cm}
\caption[Model Colours]{\label{model}An optical and near-IR colour 
diagram showing quasar model colours (described in the text).  The
colours of Galactic stars are computed from the atlas of Bruzual,
Persson, Gunn and Stryker included in the IRAF/STSDAS package SYNPHOT.
The reddening vector does not drive the quasars into the stellar locus
unlike purely optical colour-colour plots (see figure \ref{ugr}).  The
unfilled circles indicate the colours of known quasars in SDSS
(Richards \etal 2001), for which we obtained H magnitudes by looking
for matches (within 2'') in the 2MASS second incremental release point
source catalogue.}
\end{figure}
\begin{figure}
\epsfig{file=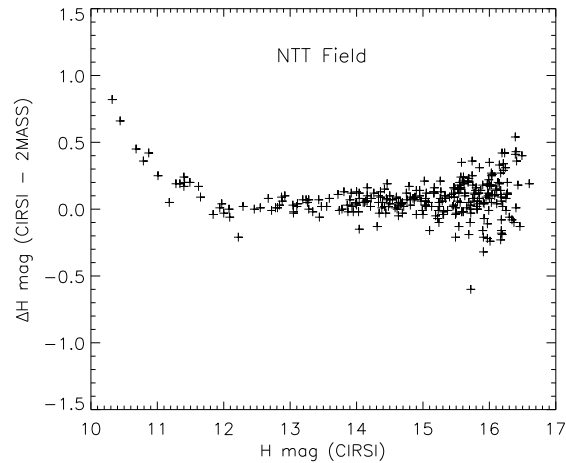,width=8cm}
\caption[Magnitude Calibration]{\label{plotdh}The magnitude difference
between CIRSI and 2MASS catalogues are shown for the 1204$-$0736 field.
The rising tail at bright magnitudes is due to saturation in the CIRSI
data.  The median magnitude offset is 0.07m.  After correcting for
this median offset, the rms scatter between the two samples is 0.12m.}
\end{figure}
\begin{figure}
\psfig{file=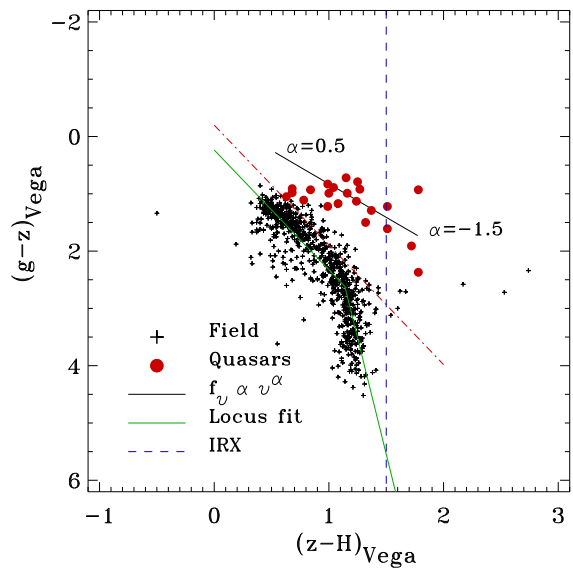,width=8cm}
\caption[Colour Diagram]{\label{plotcd}The observed optical and near-IR colour
diagram is shown for the 0218$-$0500 field (Table \ref{fields table}).
Identified quasars are indicated by filled circles.  The selection
boundary (dot-dashed line), chosen based on model quasar colours, is
$(z-H) > 0.59 \times (g-z) - 0.06$.}
\end{figure}
\begin{figure}
\epsfig{file=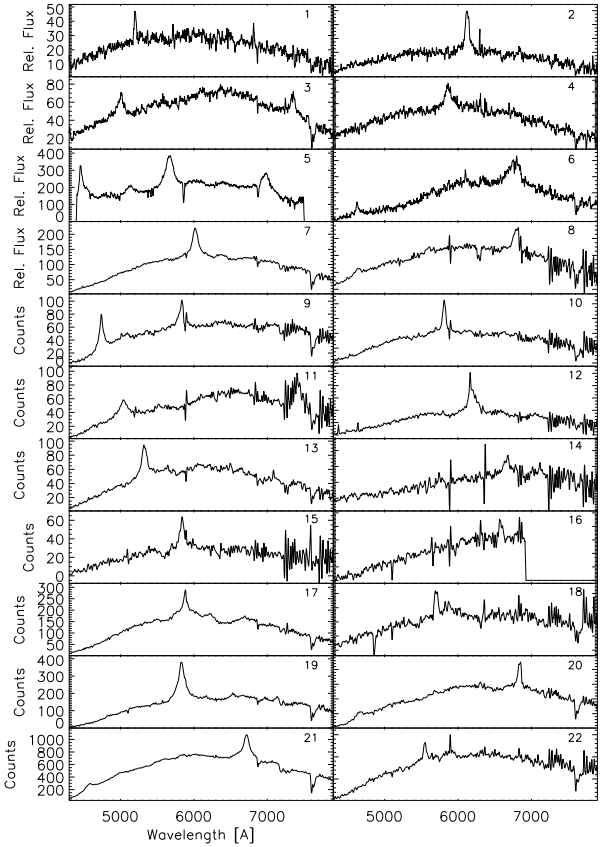,width=8cm}
\caption[Quasar Spectra]{\label{plotspc}The follow-up spectroscopy for the 
quasar sample is shown.  The first six spectra in the plot were taken
with a long-slit spectrograph on the du Pont telescope.  The remaining
spectra were taken with the Hydra multi-object spectrograph on WIYN.
The flux calibrated spectrum of object 6 shows an IRX quasar while the
fiber spectra of objects 14, 15, 16 and 19 are consistent with the IRX
photometry.}
\end{figure}
\begin{figure}
\epsfig{file=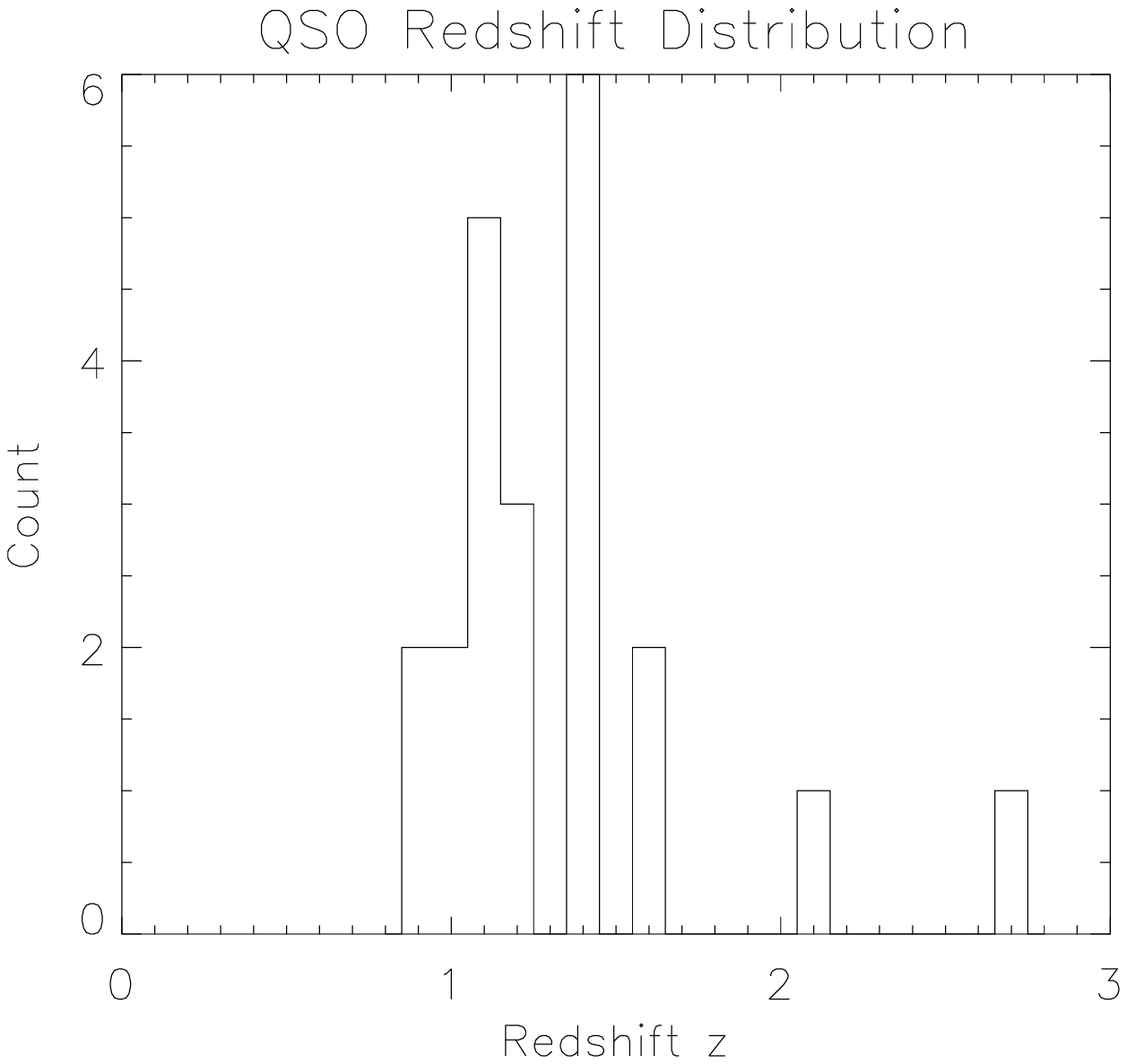,width=8cm}
\caption[Redshift distribution]{\label{plotz}The redshift distribution for the
sample of 22 confirmed quasars is shown.  The observed redshifts range
from $z = 0.858$ to $z = 2.660$ with a median redshift $z = 1.211$.}
\end{figure}
\section{Optical and Near-IR Observations}
\label{Near-IR and Optical Observations}
We use optical and near-IR imaging data for three fields covering a total
of $\approx 0.7$deg$^2$ to m$\rm_H$$\sim$18 (see Table \ref{fields table}).
These fields were chosen due to the availability of both optical and
near-IR imaging from a number of survey programs.

The optical imaging was obtained as part of the Isaac Newton Telescope
Wide Angle Survey (McMahon \etal (2001);
http://www.ast.cam.ac.uk/$\sim$wfcsur/).  This survey is being carried
out with the prime focus Wide Field Camera (WFC; Ives, Tulloch, and
Churchill 1996) at the 2.5m Isaac Newton telescope (INT) on La Palma.
The WFC consists of a closely-packed mosaic of 4 thinned EEV42
2k$\times$4k CCDs with a pixel size of 13.5$\mu$m corresponding to
0.33'' / pixel and effective field of view of 0.25deg$^2$.
Observations are taken in 5 wavebands (ugriz) with single exposures of
600sec over an area $\sim$100deg$^2$ to nominal 5$\sigma$ limiting
magnitudes of 23, 25, 24, 23 and 22 respectively.  The CCD mosaic data
is pipeline processed and calibrated in Cambridge (Irwin and Lewis
2000).  We use a preliminary photometric calibration that is accurate
to $\pm 0.1$ mag.

The near-IR imaging was obtained as part of two survey projects that
use CIRSI (the Cambridge Infrared Survey Instrument; Beckett \etal
1996, Mackay \etal 2000).  CIRSI is a JH-band mosaic imager consisting
of 4 Rockwell 1k$\times$1k
HgCdTe detectors, providing an instantaneous field of view of 15.6' x
15.6' on the INT (0.45'' / pixel) and 6.6' x 6.6' on the du Pont 2.5m
(0.19'' / pixel) at the Las Campanas Observatory.  The 1204$-$0736
field observations were taken with the du Pont 2.5m as part of the Las
Campanas IR Survey (Firth \etal 2001, Chen \etal 2001) during the
nights 29,30 December 1999; 1,2 January 2000; and 14-17 February 2000.
The 1636+4101 and 0218$-$0500 field observations were taken with the
INT on 13 June 2000 and 4,6,8 November 1998 respectively as part of
the Infra Red CIRSI-INT survey.  Total exposure times were 2000s for
the 1204$-$0736 field (up to 4300s in some regions), 2400s for the
1636+4101 field, and 2480s for the 0218$-$0500 field.  The typical
observation sequence consisted of a 7-point dither with three 30
second exposures at each dither position, and four telescope pointings
to fill in the gaps between the chips.

The CIRSI near-IR data were reduced and mosaiced using an automated
pipeline (Sabbey \etal 2001), and object catalogs were produced using
SExtractor (Bertin and Arnouts 1996).  The photometric zeropoint was
established by comparison to 2MASS (see Figure~\ref{plotdh}).  With
the precise astrometry available in both optical and infrared catalogs
(0.3'' rms with respect to the APM catalog), we were able to merge the
catalogs by looking for positional matches within 1''.  The merged
catalog was limited to point sources to remove galaxies from the
quasar candidate list.  For the spectroscopic observations undertaken
to date, the star/galaxy seperation was perfromed by requireing a
stellar classification for candidate objects in gr and i band imaging
data.  Section \ref{host} discusses the implication of this
restriction to the quasar candidate list and suggests that the
criterion should be relaxed to a g band stellar classification alone
to prevent lower redshift quasars, with resolved host galaxies, being
removed from the candidate list.
\section{Colour Selection of z$<$3 quasar candidates}
\label{colours}
Figure \ref{model} shows model quasar colours computed assuming an
underlying quasar spectrum based on a power law with spectral index
$\alpha$=$-$0.50 ($S_\nu\propto \nu^{\alpha}$) and with an emission
line spectrum based on the Francis \etal (1991) composite spectrum.
The influence of a range of values for $\alpha$ is indicated in the
plot.  The reddening vector is derived over a range of redshifts using
the galactic reddening model of Seaton (1979).  Synthetic photometry
calculations are performed with the IRAF/STSDAS package SYNPHOT.  The
stellar locus is calculated syntheticaly using the stellar atlas of
Gunn-Stryker (1983) using the extended spectral coverage spectra from
the Bruzual-Persson-Gunn-Stryker spectrophotometric atlas found in
SYNPHOT.  Representative filter transmission curves for the observing
system are applied to the input spectra to predict the colours of
objects.

\begin{figure}
\psfig{file=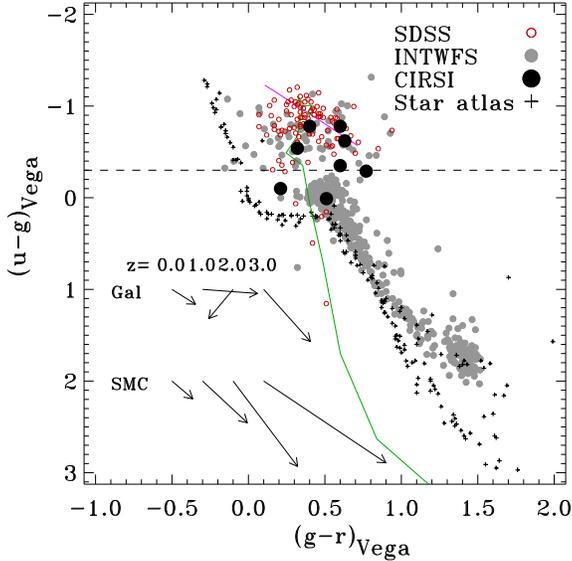,width=8cm}
\caption[]{\label{ugr}Multi wavelength data is available in ugrizJH
for a subset of the current sample.  Detailed analysis of the multi
band properties of the full quasar sample is underway.  Here we show a
comparison of a sample of previously known quasars identified in SDSS
(Richards \etal 2001) with the new quasars identified by gzH selection
for which ugr data is available.  The horizontal line at u$-$g=$-$0.3
represents a UVX selection boundary analogous to that used in the 2df
quasar survey (2QZ Boyle \etal 2000).  The stellar locus, computed
from Gunn and Stryker (1983), is shown along with the observational
data from one field ($\sim$0.25deg$^2$) from the INT WFS.  The locus
of quasar colour with redshift is indicated by the solid line.
Reddening vectors for dust models based on Galactic (Gal-Seaton 1979)
and Small Magelanic Cloud (SMC-Prevot \etal 1984) are shown over a
range of redshifts.  The 2200\AA\ feature in the galactic law passes
through the g band over the range 1$<$z$<$2.}
\end{figure}

The optical and near-IR colour diagram (figure \ref{model}) shows that
quasar candidates are expected to lie separated from the stellar locus
in colour space and that reddening due to dust, either in the host
galaxy or along the line of sight to the quasar, will not drive the
quasars into the stellar locus (in contrast to purely optical
colour-colour plots, see figure \ref{ugr}).  Reddening of foreground
stars moves stars parallel to the stellar locus preventing a broadened
stellar locus overlapping the quasar region.

The mainly optically selected quasar sample of Richards \etal (2001)
with SDSS photometry and matched with the 2MASS second incremental
release is overlayed for comparison.  There is evidence that the
colours of the three quasars hidden with in the stellar locus in this
plot are strongly effected by variability (NED identifiers
NGC0450:0126$-$0019 (z=1.76), UM357:0140$-$0050 (z=0.33),
PKS1215$-$002 z=(0.42)).  One object, UM357:0140$-$0050, is a known
Optical Violent Variable (OVV) and all three objects are located on
the quasar power law spectrum locus in g$-$z v J$-$K colour space.
This latter space is insensitive to variability since for 2MASS the IR
data is recorded simultaneously through the use of dichroics and for
the SDSS bands the lag between observation is insignificant
($\sim$minutes) when compared to the variability timescale for broad
band optical observations of quasars.

To identify quasar candidated the stellar locus is approximated by a
two component linear fit of the form,
\begin{eqnarray}
(z-H) &>& m \times (g-z) + c + offset
\end{eqnarray}
The offset value being chosen to minimize the number of stray stars in
the sample while rejecting a minimum of quasar candidates.

The region above these boundaries (lower g$-$z for a give value of
z$-$H) defines the candidate selection region.  All stellar objects in
this region are visually inspected to check for spurious photometry
due for example to stellar diffraction spikes or cosmic rays in the
single exposure INT WFS data.

At redshift z$>$3, the Lyman-$\alpha$ forest passes into the g band
and absorption in the Inter Galactic Medium (IGM) moves quasar colours
into the stellar locus.  Few z$>$3 quasars would therefore be expected
in a gzH selected sample.
\subsection{Prospects for extending the technique to z$>$3.0}
\begin{figure*}
\centering
\psfig{file=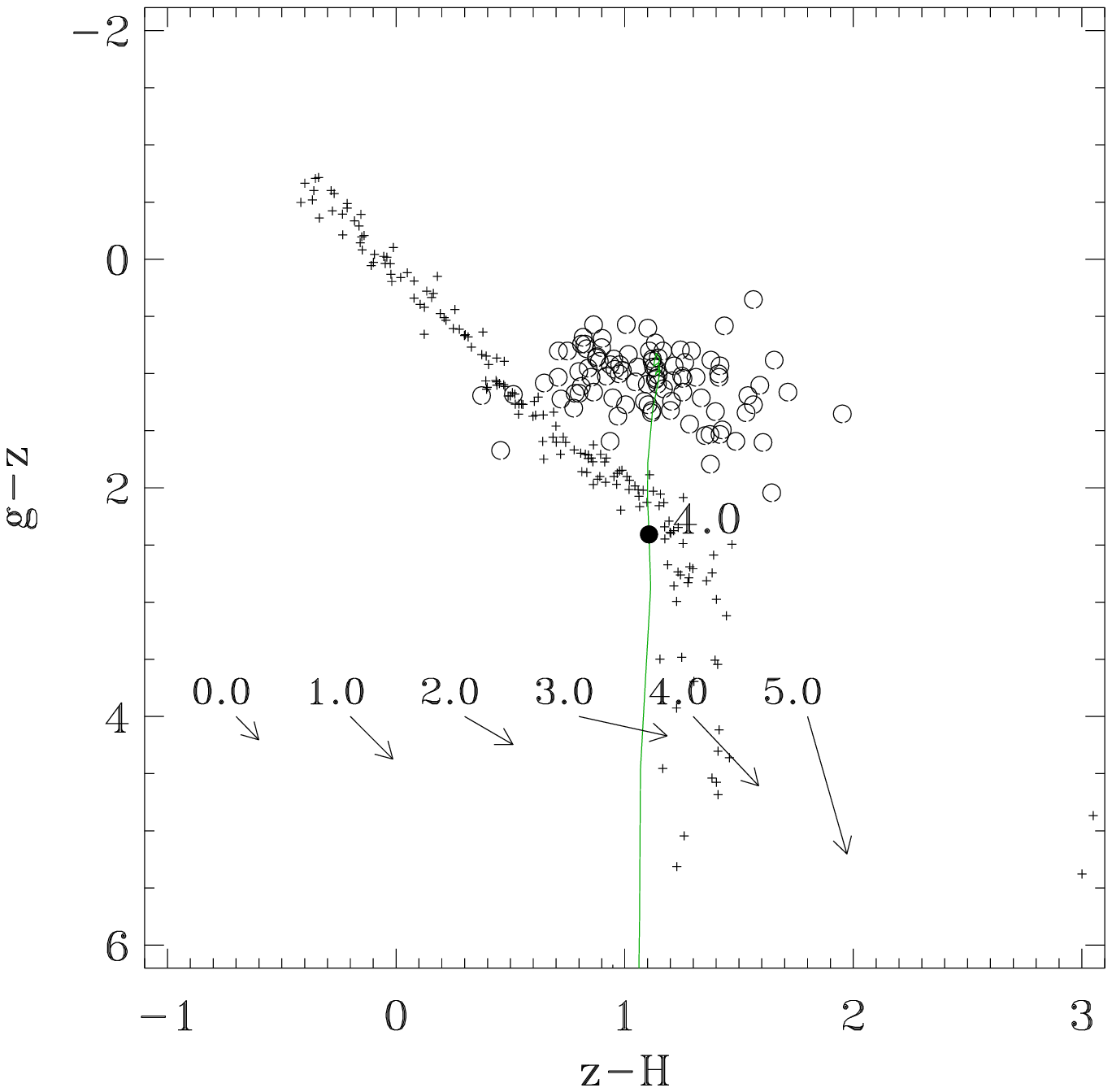,width=7cm}
\psfig{file=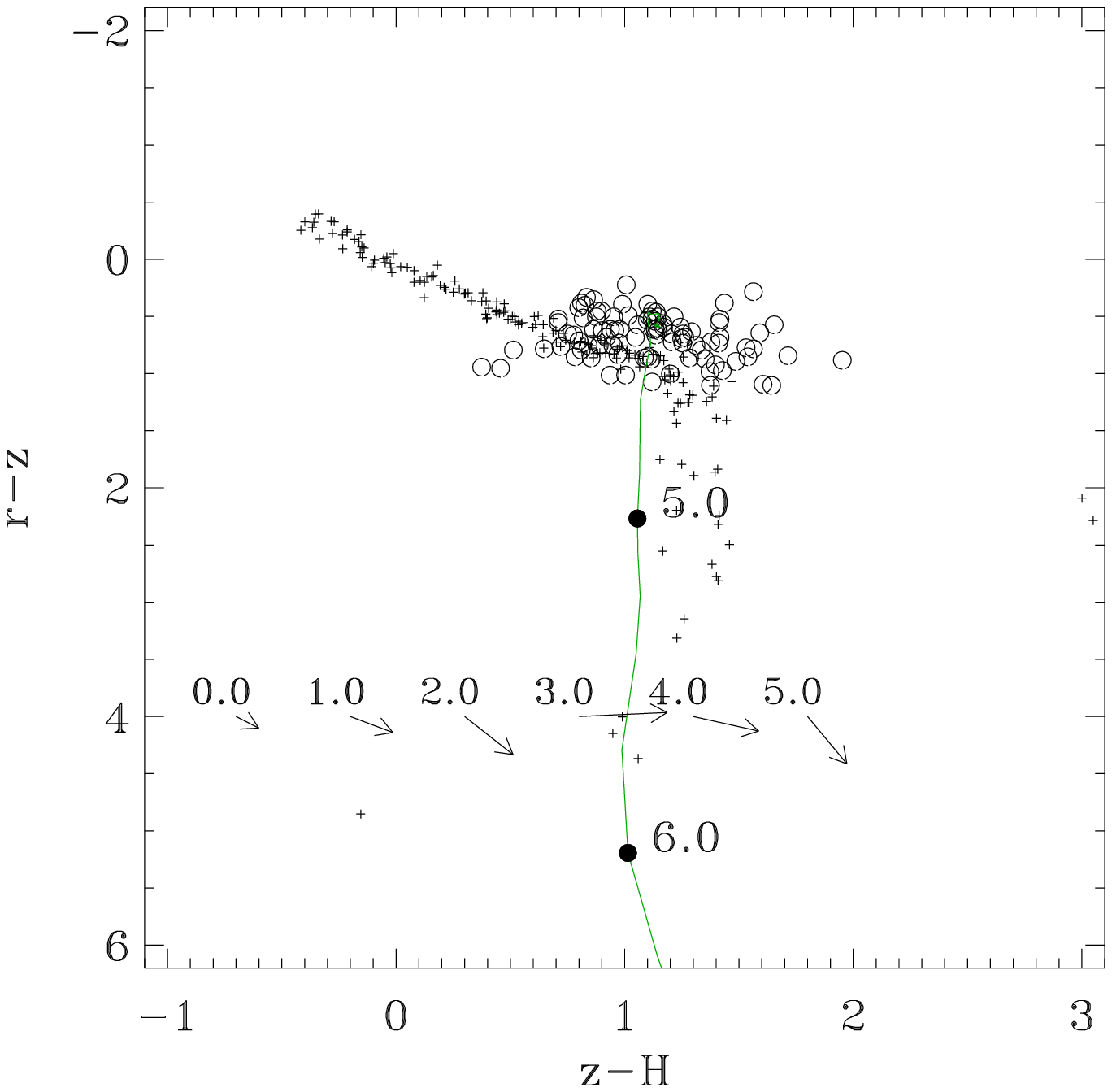,width=7cm}
\psfig{file=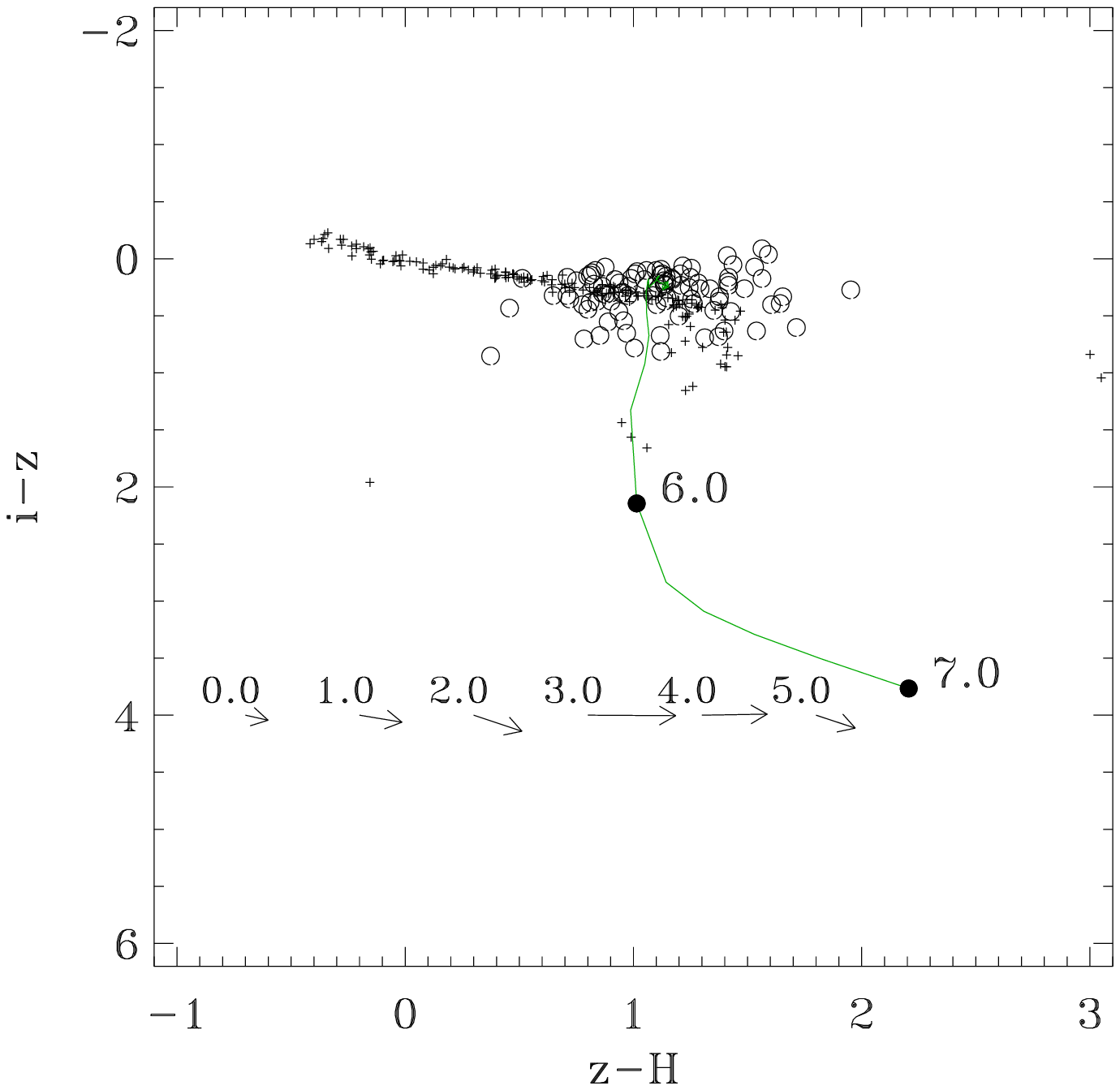,width=7cm}
\psfig{file=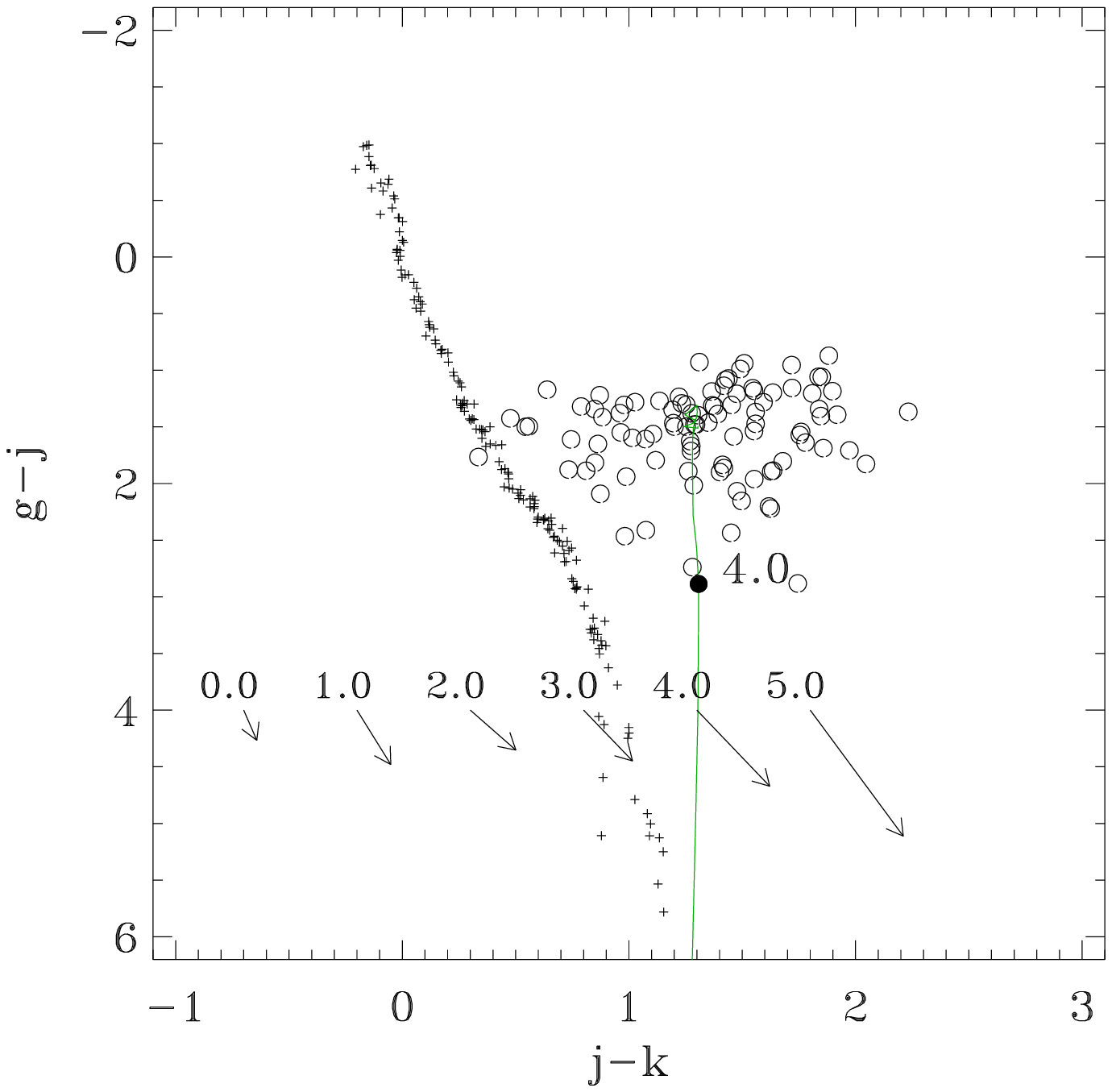,width=7cm}
\psfig{file=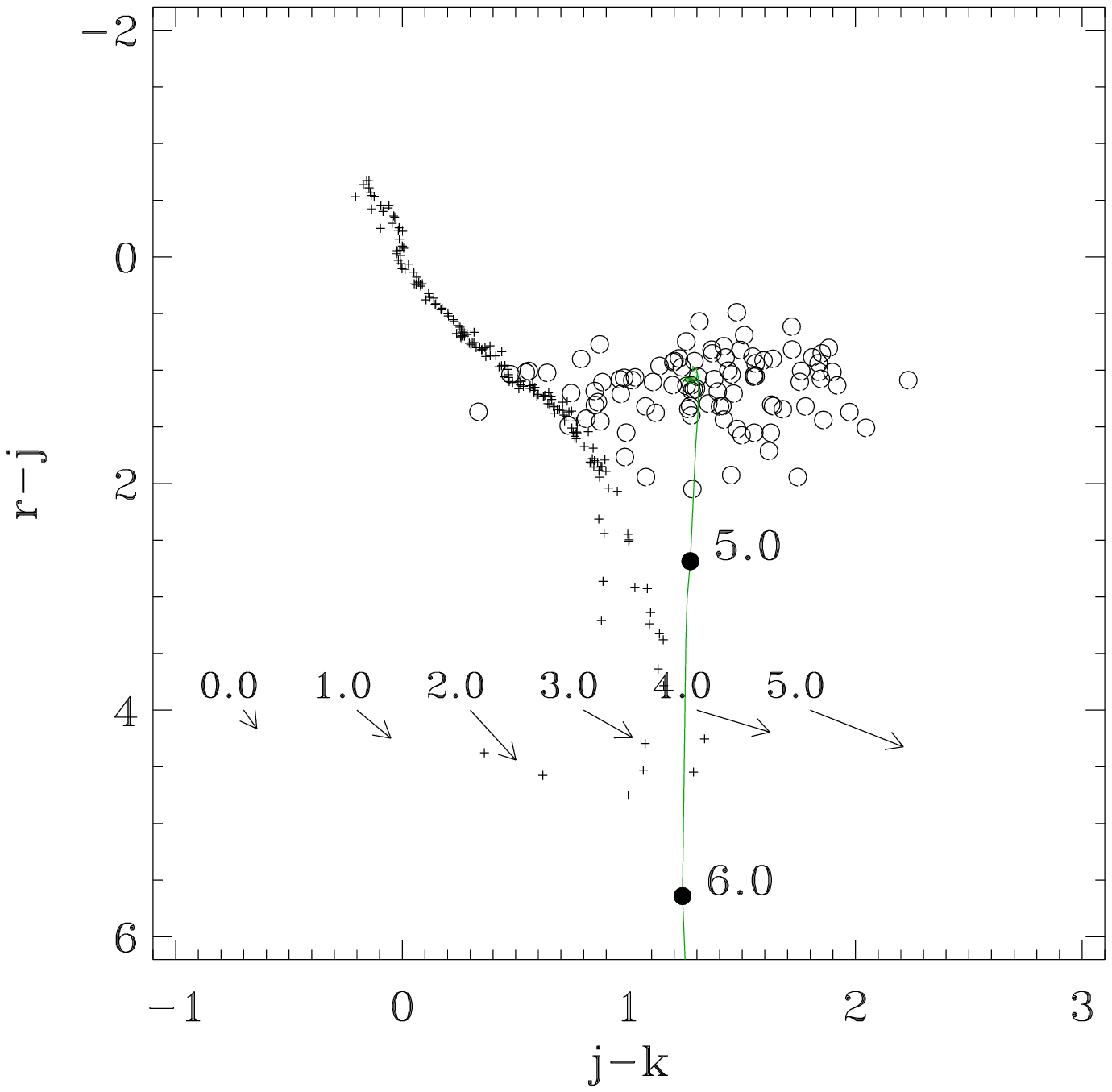,width=7cm}
\psfig{file=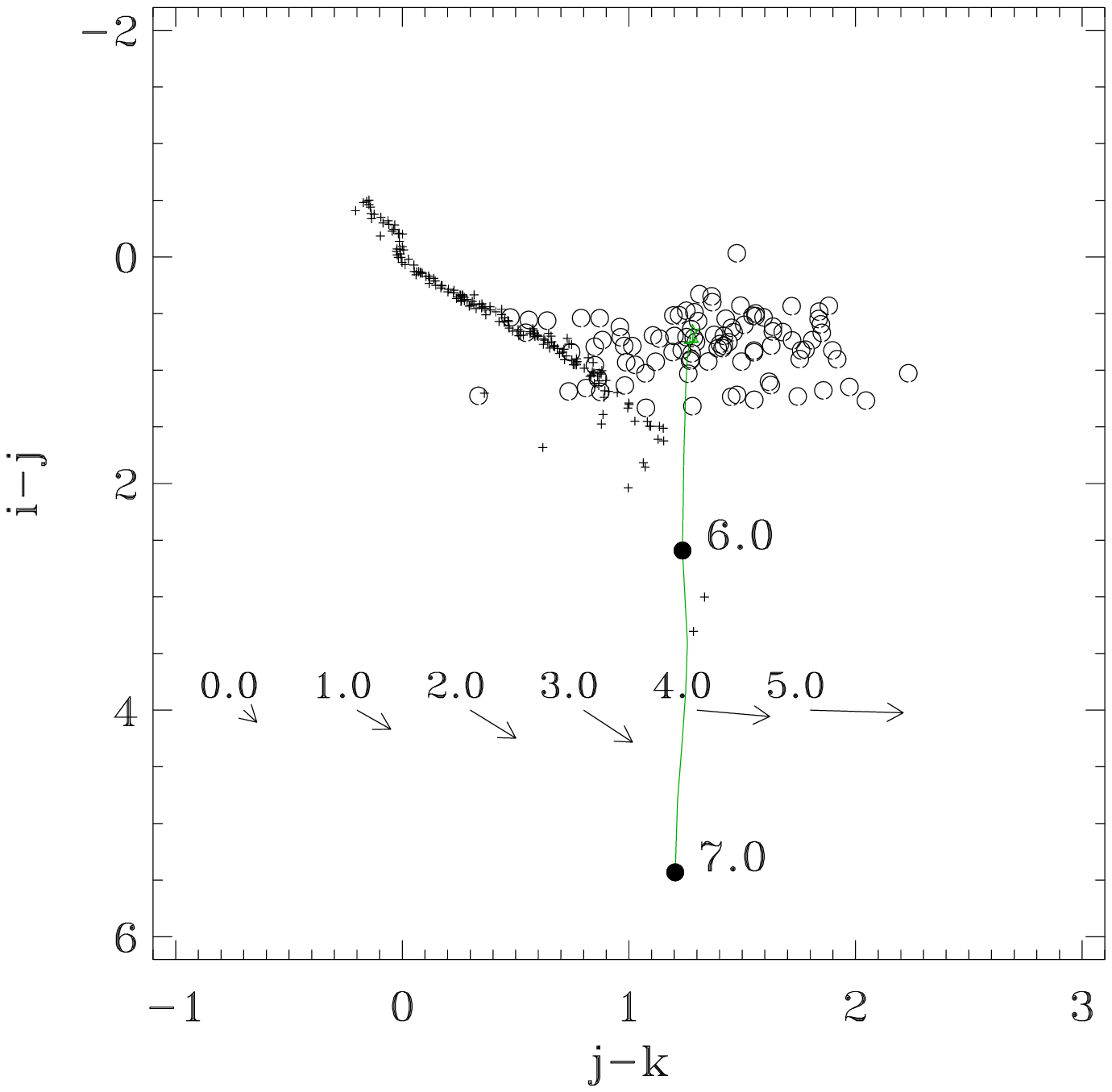,width=7cm}
\caption[]{\label{lots of plots}A range of potential optical-IR quasar
colour selection diagrams are considered.  A sample of quasars is
constructed from the SDSS quasar sample of Richards \etal (2001) with
IR data obtained from a 2'' radius match to the 2MASS second
incremental release data.  A stellar locus is computed from the
spectral atlas of Gunn and Stryker (1983).  Reddening vectors are
shown for A$\rm_V$(rest)=1.0 under the Seaton (1979) galactic
reddening law.  The quasar locus with redshift is marked for a range
of redshifts.}
\end{figure*}

As part of the Isaac Newton Telescope Wide Angle Survey (Walton,
Lennon, Irwin and McMahon 2001) optical observations, over the full
range of wavelengths ugri and z, are available for two of the fields
examined in this paper.  This data set has been successfully used for
the identification of high redshift (z$\sim$5) quasars via optical
colour selection techniques (Sharp, McMahon, Irwin and Hodgkin 2001).
However, the gzH technique can not be used at higher redshift (z$>$3,
after the onset of IGM absorption in the g band) by replacement of the
g band alone with the r or i bands as quasars merge with the stellar
locus (as shown in the colour diagrams of figure \ref{lots of plots}).
Figure \ref{lots of plots} shows, as discussed by Barkhouse and Hall
(2001), it is possible to use the RJK diagram.
Pushing the technique into the i band and using the iJK diagram allows
the selection of targets out to redshift 5 (before IGM absorption
confuses the selection with the low mass region of the stellar locus)
but blending with the main sequence is greater than that in the rJK
diagram.  Figure \ref{lots of plots} demonstrates a range of possible
selection plots.

The lower surface density of 3$<$z$<$4 quasars, in comparison to the
high selection efficiency (65\%) gzH z$<$3 quasars, implies that a rzH
selected quasar sample could be effectively integrated into a joint IR
quasar selection campaign, using wide field of view multi-object
spectrograph observations,
capitalizing on the high multiplex capabilities of instruments such as
HYDRA.
\section{Potential effects of the host galaxy}
\label{host}
\begin{figure}
\psfig{file=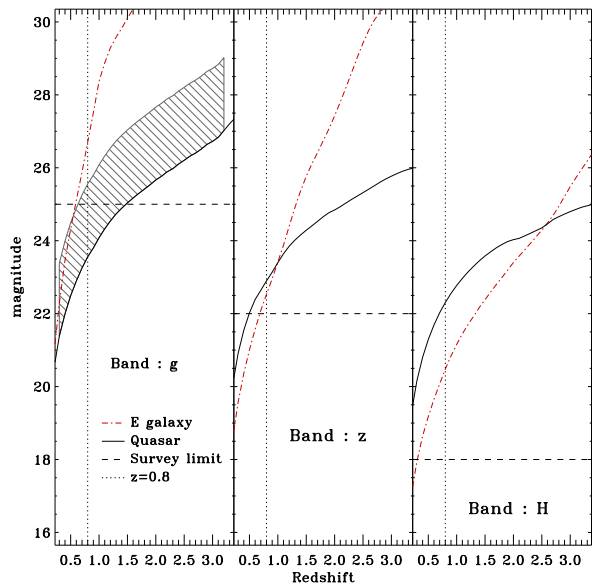,width=8cm}
\caption[]{\label{host galaxies plot}
The magnitude evolution tracks are shown for an early type galaxy and
the model quasar spectrum.  Both objects are normalized to
M$^*$=M$\rm_{Bj}$=$-$19.5 (Madgwick \etal 2001) for the comparison.
Figure \ref{host galaxy MB} demonstrates the observed range in quasar
and host galaxy magnitudes.  The galaxy template used is taken from
Coleman, Wu and Weedman 1980 and no evolution in the spectrum is
invoked.  The nominal survey 5$\sigma$ limits are marked by the dashed
horizontal line.  While the galaxy contribution to the g band is weak
it may dominate the flux in the IR H band.  The shaded area indicates
a 2 magnitude wide zone below the PSF limited quasar magnitude.  If
the host galaxy aperture magnitude lines within this zone an object
may be classified as an extended source.}
\end{figure}
\begin{figure}
\psfig{file=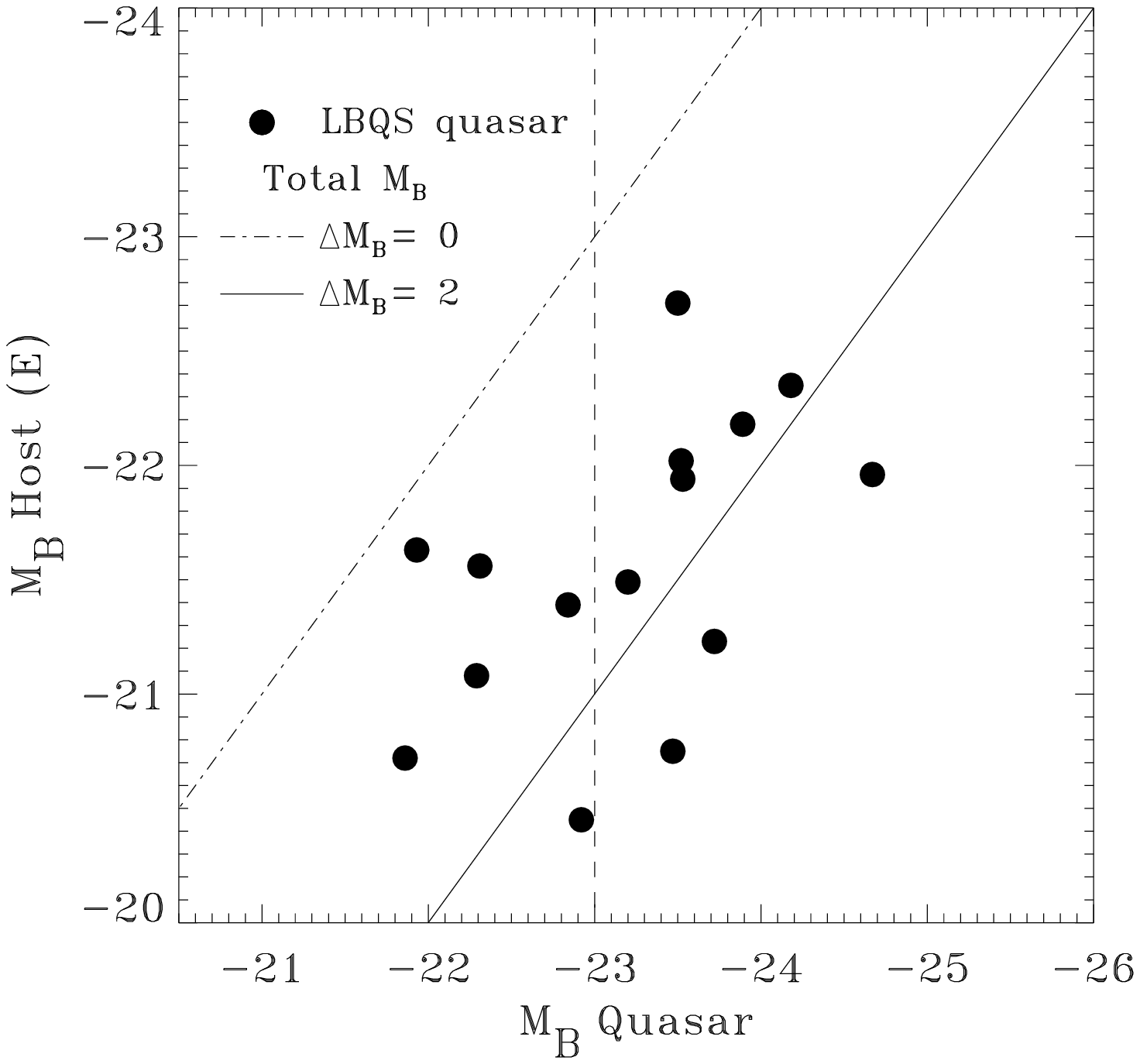,width=8cm}
\caption[]{\label{host galaxy MB} A comparison of M$\rm_B$ for a
sample of quasars, taken from the LBQS sample and assuming elliptical
host galaxies, is obtained from Hooper, Impey and Foltz (1997).
Conversion from M$\rm_R$ to M$\rm_B$ is performed using the quasar
model described in the text and the galaxy template of Coleman, Wu and
Weedman 1980.  Values for M$\rm_R$ where derived from profile fitting
and represent total magnitudes.}
\end{figure}
There are two potential problems associate with the detection of the
quasar host galaxy. Both are illustrated in figure \ref{host galaxies plot}.

1) The magnitude limit of the gzH selection technique is essentially
 set by the depth of the H band data due to the relative easy with
 which deep g and z observations can be achieved.  However the depth
 of the INT WFS imaging data is sufficient that potentially the host
 galaxy of quasars may be detected and spatially resolved in one or
 more of the optical passbands.  Star/galaxy separation is used to
 remove many galaxies from the sample and this step may also reject
 lower luminosity quasars with bright host galaxies.

2) The addition of a galaxy component to the observed flux may alter
   the objects colour to the extent that it would no longer fall
   within the colour selection boundary.

The absence of any quasar with z$<$0.8 in our current sample is
indicative of the first problem.  Using the luminosity function of
Boyle \etal (2000), and within the uncertainties of the conversion
from m$\rm_H$ to M$\rm_B$, we expect $\sim$2.6 quasars per deg$^2$ to
m$\rm_H$$<$18.0 in the range 0.0$<$z$<$0.8 and M$\rm_B$$<$$-$23.
However with only $\sim$47\% of candidates observed at this time and a
nominal survey area of $\sim$0.7deg$^2$ this becomes $\sim$0.9 quasars
that may be missing from the sample, not inconsistent with the small
sample size.

Above z=0.8 the optical g band, with which star/galaxy
separation is performed, samples the rest frame UV continuum (longword
of the 4000\AA\ break) where the host galaxy presence in imaging data
will be at it's weakest.

Morphological classification of objects within the INT WFS is based on
an analysis of the curve of growth of an objects flux in apertures of
radius 1/2, 1 ,$\sqrt{2}$, 2, 2$\sqrt{2}$ times the seeing for the
observation.  A detailed discussion of the classification algorithm is
beyond the scope of this paper.  However investigation of the aperture
flux ratio and classification of simulated and observed early type
galaxies and PSF limited point source profiles show that quasar cores
will be classified as stellar in the INT WFS g band in the presence of
an extended host galaxy component provided the host is 2 magnitudes
fainter than the quasar core (assuming a De Vaucouleur light profile
for the host galaxy with half-light radius 0.5'' and axis ratio 0.8).
This comparison is based on the seeing radius aperture magnitude as
oppose to a total magnitude.  A stellar classification will be
obtained for the quasar core of a composite object for a lower
separation in total magnitude, the value depending strongly on the
definition of total magnitude.

\begin{figure}
\epsfig{file=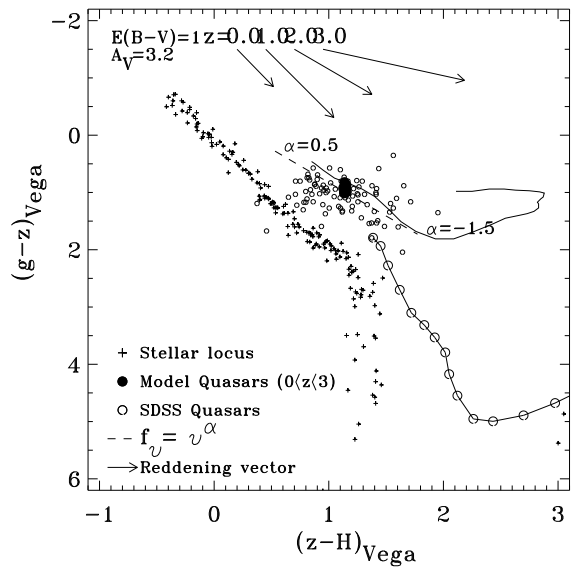,width=8cm}
\caption[Galaxy colour tracks]{\label{galtrack}The gzh diagram from
figure \ref{model} is shown overlayed with the colour evolution track
for a non evolving early type galaxy (0$<$z$<$3, 0.1 z step).  The
colour track for a combination of quasar and elliptical galaxy
spectrum is shown with the elliptical component normalized the 4
magnitudes below the quasar in the rest frame B band.  The galaxy
component dominates the H band flux.}
\end{figure}

Figure \ref{galtrack} addresses the second effect.  A reproduction of
figure \ref{model} is shown with the colour evolution track
(0$<$z$<3$, 0.1 steps in z) for an non evolving Early type galaxy
overlayed.  A colour evolution track assuming a combination of the
model quasar and such a host galaxy, with the galaxy normalized to 4
magnitudes fainter than the quasar in the rest frame B band, is
indicated by the solid line.

While it is clear that any host galaxy contribution to the quasar
colour has a significant effect, it acts to increase the detectability
of the quasar in the sense that it improves separation from the
stellar locus.

From Figure \ref{host galaxies plot}, the nominal H band magnitude
limit of our survey of m$\rm_H$=18.0 corresponds to an absolute B band
magnitude of $-$24.5 and $-$25.5 at z=1 and 2 respectively. Therefore
if the quasar host galaxy is uncorrelated with the quasar luminosity
and is assumed to correspond to an unevolved L* ellipical galaxy
(M$\rm_B$=$-$19.5, Madgwick \etal 2001) the quasar host will be 3-5
magnitudes fainter than the quasar nucleus. However, if the quasar
host galaxy is correlated with the quasar luminosity such as one would
expect from the observed correlations in the local Universe between
dormant quiescent Black hole masses and galaxy bulge luminosities
(Magorrian \etal 1998) the quasar host galaxy may contribute to the H
band light. Taking the observed correlation between rest frame
absolute magnitudes of z$\sim$0.5 quasars and quasar hosts observed by
Hooper, Impey and Foltz (1997) where $\Delta$M$\rm_B$=2 (see figure
\ref{host galaxy MB}) the quasar host may contribute $\sim$50\% of the
light in H in the redshift range 1 to 2 (see figure \ref{host galaxies
plot}).

\section{Spectroscopic Observations}
\label{spectra}
Multi-object spectroscopy was obtained using the Hydra multi-fiber
spectrograph (Barden and Armandroff 1995) on the 3.5m WIYN telescope
at the Kitt Peak Observatory.  The 1637+4101 and 0218$-$0500 fields
were observed during 2 and 3 September 2000 (as part of an observing
program containing other projects) in clear sky conditions with a
seeing of $\approx 0.8$''.  We used the Bench Spectrograph Camera and
a 316 lines/mm grating with a central wavelength of 6027\AA. The total
wavelength coverage was from 4240\AA\ to 8728\AA\ (the blue end was
set by a long pass filter). The dispersion was 2.64\AA/pix and the
resolution was 5.7\AA.  Arc lamp (Cu-Ar) calibration frames were taken
to set the wavelength scale.  With the 1 degree field diameter and 97
red fibers (2'' diameter) available, we were able to obtain spectra
for both quasar candidates and a sample of random targets spread
throughout the colour diagram.  The minimum exposure time was 1.7
hours, although many objects were observed twice in overlapping Hydra
fields resulting in 3.4 hour exposures.

Long-slit spectroscopic observations were taken during the five nights
from 22 to 26 February 2001 at the du Pont 2.5m telescope at Las
Campanas Observatory.  The weather conditions and seeing were both
good.  The Modular Spectrograph was used with the SITe\#2 detector and
the 150mm grating to obtain spectra with 4\AA/pixel over the
wavelength range 4300-8500\AA.  We obtained spectra for 38 objects
(including separate projects), with exposure times ranging from 10
minutes to 1 hour. Wavelength calibration was performed with arc
spectra.
\section{Confirmed Quasars}
\label{quasars}
\begin{figure}
\psfig{file=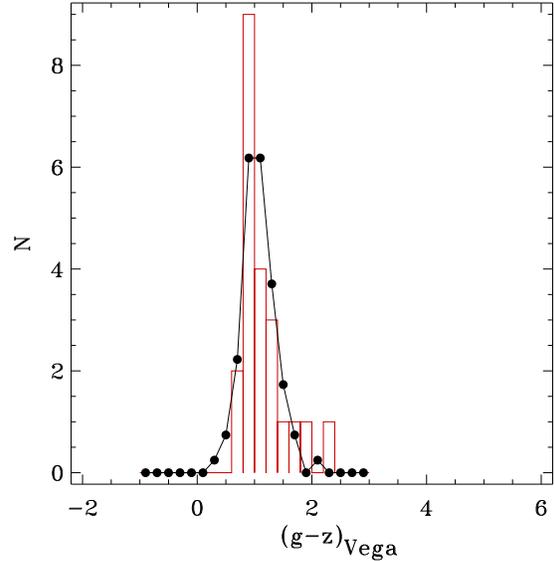,width=8cm}
\psfig{file=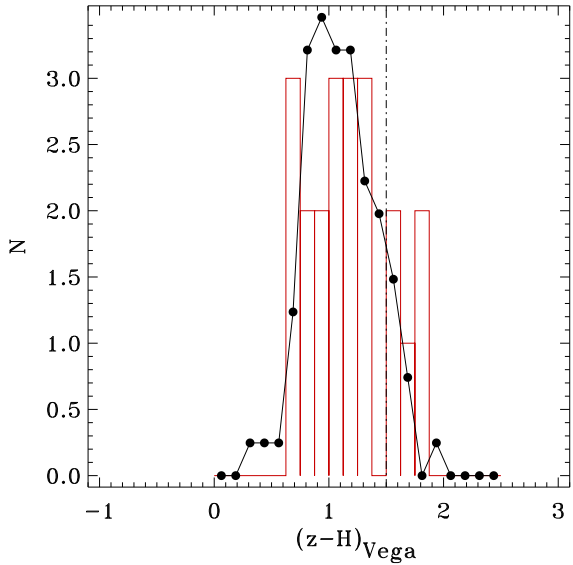,width=8cm}
\caption[]{\label{colour hist}A comparison of the histograms of
quasars colours for the CIRSI sample and that of Richards \etal
(2001), from SDSS and matched to the 2MASS point source catalogue,
shows broad similarity between the distributions.  The SDSS sample (89
quasars with z$<$2.6) is scaled to the CIRSI sample size of 22
quasars.  A K-S test between the distributions yields probabilities
that they of are drawn from the same parent population of :
g$-$z=0.82, z$-$H=0.76.}
\end{figure}
The confirmed quasars are listed in Table \ref{quasars table}.  Object
16 has a possible broad emission-line at 6500\AA, but the spectrum is
low signal-to-noise and requires independent confirmation.  Based on
Veron-Cetty and Veron (2000), 2 of the 22 confirmed quasars were
previously known.  Both known quasars where discovered as part of the
slitless spectroscopy survey of Crampton \etal (1988).  Three
additional objects are listed in the NED data base as candidate
quasars from Crampton, Cowley, Hartwick and Ko (1992) but without
prior spectroscopic confirmation and redshifts.  The redshift
distribution is shown in figure \ref{plotz}.  The median redshift is
z=1.211.

The current spectroscopic sample size hampers a full comparison of the
quasar space density with other works.  The sample is m$\rm_H$ limited
making comparisons with optical samples complex due to uncertainties
in the K-correction required to compute optical absolute magnitude
indicators.  The sample is not complete to an optical magnitude limit
in m$\rm_g$ or m$\rm_z$ due to incomplete follow up of the fainter
optical candidates.

U band observations are available for a subsample of 8 quasars.  Two
quasars, 17 and 22, do not exhibit strong UV emission and could be
missing from a classical UVX selected sample.  Defining a IRX
criterion of z-H$>$1.5, neither object is IRX (z-H=1.27 and 1.37
respectively). However, five IRX quasars are identified in the sample,
quasars 6, 14, 15, 16 and 19.  Objects 15 and 16 show low UVX (-0.35
and -0.30) while quasar 19 has a more typical value of -0.6.  No u
band observations are available for objects 6 and 14 at this time.

Figure \ref{colour hist} compares the histogram of quasar colours for
the 22 quasar CIRSI sample and 89 redshift z$<$2.6 quasars from the
Richards \etal (2001) sample matched to the 2MASS point source
catalogue to provide H band magnitudes.  The Richards \etal (2001)
sample is normalized to the total sample size of the CIRSI
observations.  Within the limited sample size the colour distributions
are similar with a possible excess of IRX quasars in the CIRSI
sample.\\

As discussed in section \ref{host}, the lack of quasar in the sample
with redshift z$<$0.8 is not inconsistent with the luminosity function
predictions within the constraints of the current survey area
(0.7\sqdeg) and $\sim$47\% candidate follow up.  The low number of
quasars at z$>$2 is also not inconsistent.
\section{Conclusions}
We report the identification of a gzH selected sample of z$<$3
quasars.  The $g-z/z-H$ colour diagram is used with high selection
efficiency ($\sim$65\%) and reduced sensitivity to reddening and
extinction.  The initial sample has 22 confirmed quasars, but the full
INT CIRSI survey will contain several hundred.  To date two objects
within a subsample of the identified quasars, for which UV
observations are available, do not possess a significant UV excess and
would not have been recovered by a UVX based selection.  The full
CIRSI-INT gzH quasar sample will cover an area of in excess of
10deg$\rm^2$ \ including observations across the full range of bands
ugrizJ and H.  Contrasting the sample with those compiled through a
range of selection techniques, such as UVX colour selection and radio
identifications from overlapping surveys such as the VLA-FIRST survey,
will provide an excellent tool for identifying biasing within quasar
selection methodologies.

\begin{table*}
\caption{\label{fields table}Survey fields}
\begin{center}
\begin{tabular}{ccccccc}
Name & \multicolumn{2}{c}{R.A. (J2000) Dec} & Area &
     \multicolumn{3}{c}{Depth (Vega)} \\ & & & deg$^2$ & $g$ & $z$ &
     $H$ \\
\hline
1204$-$0736 & 12:04:50   &  $-$07:36:00 & 0.18  &  24  & 21  & 18 \\
1636+4101 & 16:36:50     &  +41:01:50 & 0.35 & 24 & 21 & 18 \\
0218$-$0500 & 02:18:00   &  $-$05:00:00 & 0.17 & 24 & 21 & 17.5 \\
\hline
\end{tabular}
\end{center}
\end{table*}
\begin{table*}
\label{quasars table}
\caption{The quasar Sample.}
\begin{center}
\begin{tabular}{cccccccccc}
Name & \multicolumn{2}{c}{R.A. (J2000) Dec} &
           \multicolumn{3}{c}{magnitude (Vega)} & $z$ & g-z & z-H &
           Comments\\ & & &$m_{g}$ & $m_{z}$ & $m_{H}$\\
\hline
CIRSI1 & 12:03:48.61 & $-$07:19:00.7 & 19.98 & 18.76 & 17.77 & 0.858               & 1.22 & 0.99 \\ 
CIRSI2 & 12:04:21.62 & $-$07:22:01.0 & 19.46 & 18.41 & 17.78 & 1.189               & 1.05 & 0.63 \\ 
CIRSI3 & 12:04:09.20 & $-$07:24:22.8 & 19.89 & 18.96 & 18.12 & 1.625               & 0.93 & 0.84 \\ 
CIRSI4 & 12:04:08.18 & $-$07:21:45.6 & 20.13 & 19.41 & 18.26 & 1.097               & 0.72 & 1.15 \\ 
CIRSI5 & 12:05:34.59 & $-$07:41:21.4 & 19.08 & 17.97 & 17.19 & 2.660               & 1.11 & 0.78 \\ 
CIRSI6 & 12:05:07.68 & $-$07:45:26.6 & 21.01 & 19.40 & 17.89 & 1.420               & 1.61 & 1.51 & IRX, No UV data\\ 
CCS88 163351.6+410628 & 16:35:31.05 & +41:00:27.2 & 19.14 & 18.25 & 17.21 & 1.151  & 0.89 & 1.04 & Crampton \etal (1988)\\ 
CIRSI8 & 16:37:10.03 & +40:56:42.9 & 20.21 & 19.22 & 18.06 & 1.434                 & 0.99 & 1.16 & Candidate Crampton \etal (1992)\\ 
CIRSI9 & 16:37:00.64 & +41:05:55.2 & 19.69 & 18.78 & 18.10 & 2.061                 & 0.91 & 0.68 & Candidate Crampton \etal (1992)\\ 
CIRSI10 & 16:36:47.16 & +41:03:35.0 & 20.54 & 19.37 & 18.29 & 1.077                & 1.17 & 1.08 \\ 
CIRSI11 & 16:37:14.80 & +41:12:32.6 & 20.16 & 19.18 & 18.50 & 1.642                & 0.98 & 0.68 & Candidate Crampton \etal (1992)\\ 
CIRSI12 & 16:35:30.51 & +41:10:41.6 & 20.86 & 19.36 & 18.04 & 1.211                & 1.50 & 1.32 \\ 
CCS88 163447.3+405448 & 16:36:27.11 & +40:48:48.9 & 19.72 & 18.89 & 17.90 & 0.904  & 0.83 & 0.99 & Crampton \etal (1988)\\ 
CIRSI14 & 16:37:34.03 & +41:16:09.0 & 22.42 & 20.05 & 18.27 & 1.384                & 2.37 & 1.78 & IRX, No UV data\\ 
CIRSI15 & 02:17:56.51 & $-$05:06:51.7 & 20.72 & 19.79 & 18.01 & 1.087              & 0.93 & 1.78 & IRX, low UVX\\ 
CIRSI16 & 02:18:34.44 & $-$05:13:56.9 & 20.95 & 19.04 & 17.32 & 1.351              & 1.91 & 1.72 & IRX, low UVX\\ 
CIRSI17 & 02:17:11.99 & $-$04:46:19.8 & 18.92 & 18.00 & 16.73 & 1.102              & 0.92 & 1.27 & Non UVX\\ 
CIRSI18 & 02:18:08.56 & $-$05:12:23.9 & 20.22 & 19.43 & 18.18 & 1.038              & 0.79 & 1.25 \\ 
CIRSI19 & 02:18:17.42 & $-$04:51:12.5 & 19.41 & 18.19 & 16.68 & 1.085              & 1.22 & 1.51 & IRX \\ 
CIRSI20 & 02:18:13.94 & $-$04:52:41.1 & 19.80 & 18.67 & 17.43 & 1.443              & 1.13 & 1.24 \\ 
CIRSI21 & 02:18:30.57 & $-$04:56:22.7 & 17.60 & 16.61 & 15.61 & 1.401              & 0.99 & 1.00 \\ 
CIRSI22 & 02:17:21.66 & $-$05:06:28.9 & 19.97 & 18.68 & 17.31 & 0.983              & 1.29 & 1.37 & Non UVX\\
\hline
\end{tabular}
\end{center}
\end{table*}
\begin{figure}
\centering
\psfig{file=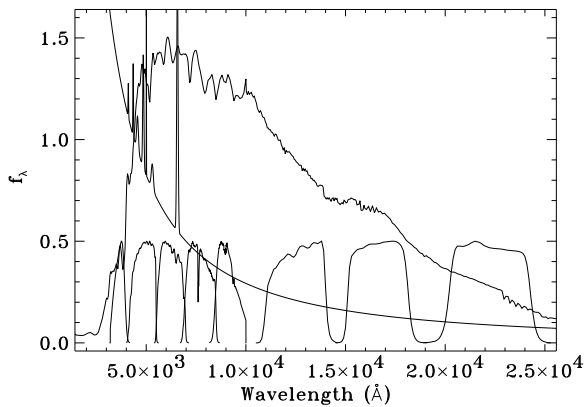,width=8cm}
\caption[Comparison of a quasar and elliptical galaxy]{\label{qso and
E}The redshift z=0 quasar and elliptical galaxy spectra used in the
discussion are shown against the rest frame filter pass bands.  Both
SEDS are normalized to 1 at $\lambda$4400\AA\ , the effective
wavelength of the B bands.  The optical ugriz and IR JHK filter sets
are shown for comparison.  Quasar - $S_\nu\propto \nu^{\alpha}$
($\alpha$=$-$0.50) with an emission line spectrum based on the Francis
\etal (1991) composite spectrum , Elliptical galaxy - Coleman, Wu and
Weedman (1980).}
\end{figure}
\begin{figure}
\psfig{file=cirsi_qso_gzh.ps,width=7cm}
\psfig{file=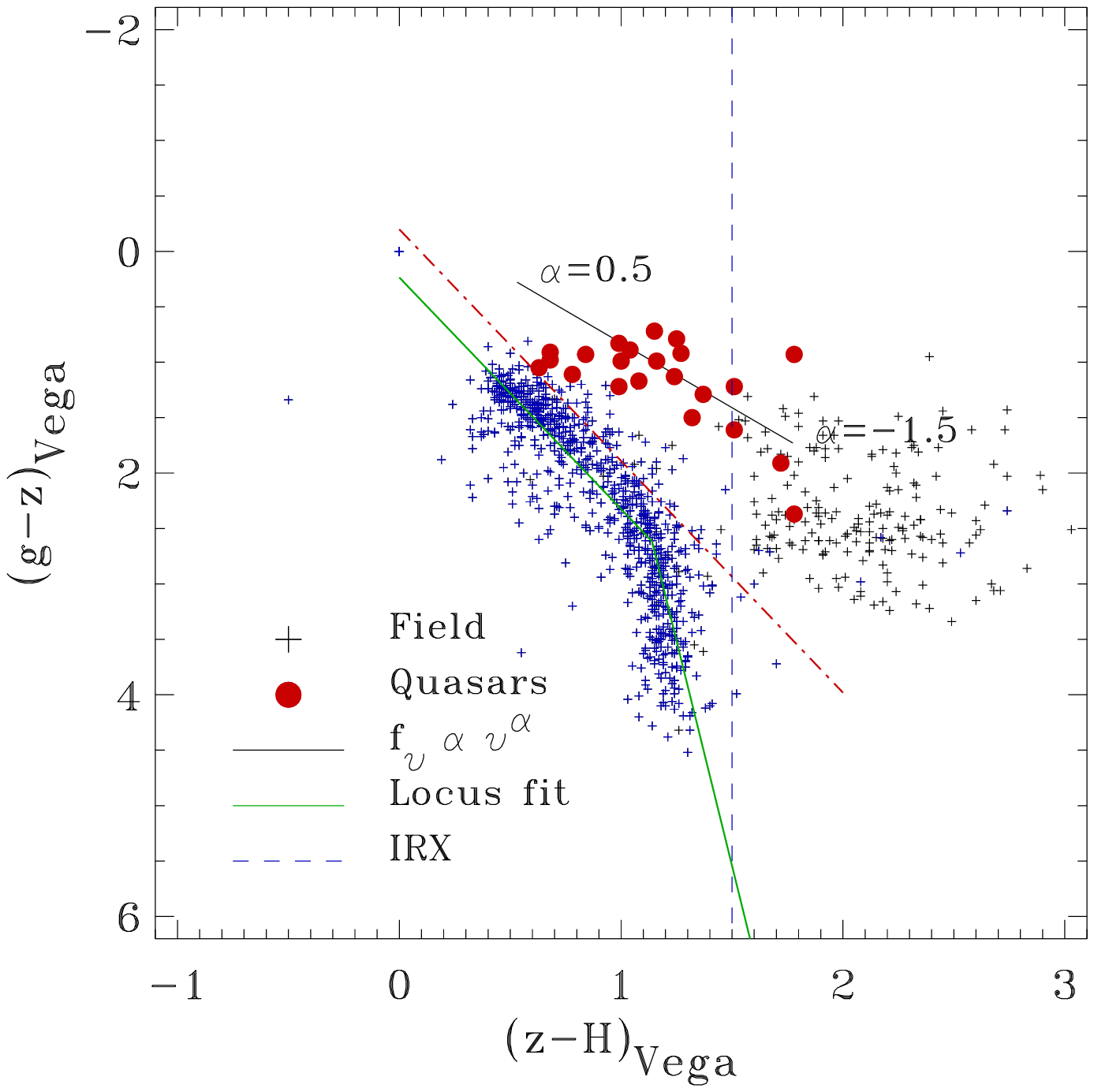,width=7cm}
\caption[]{\label{no stargal}
Figure \ref{plotcd} is reproduced to aid comparison with the identical
colour selection diagram but including unresolved and extended
objects.
Objects with profiles indicative of cosmic rays or image defects are
removed.  A limiting magnitude of m$\rm_g$=21.0 and m$\rm_H$=18.0 is
applied, based on the current spectroscopic survey.  At fainter
limiting magnitudes the growth in the number of extended sources and
of the region of colour space they occupy would make sample selection
without star/galaxy separation prohibitively expensive.}
\end{figure}
\section*{Acknowledgments}
RGS acknowledges the receipt of a PPARC Studentship. RGM thanks the
Royal Society for support.

Optical imaging data was made publicly available through the Isaac
Newton Groups' Wide Field Camera Survey Program. The Isaac Newton
Telescope is operated on the island of La Palma by the Isaac Newton
Group in the Spanish Observatorio del Roque de los Muchachos of the
Instituto de Astrofisica de Canarias.

The construction of CIRSI was made possible by a generous grant from
the Raymond and Beverly Sackler Foundation.
\section*{Bibliography}
Barden, S.C., and Armandroff, T. 1995, SPIE, 2476, 56 \\
Barkhouse,W.A., and Hall, P.B. 2001 AJ 121 2843\\
Beckett, M.G. \etal 1996, SPIE, 2871, 1152 \\
Bertin, E., and Arnouts, S. 1996, A\&AS, 117, 393\\
Beichman C.A., Chester T.J., Cutri R., Lonsdale C.J., Kirkpatrick D., Smith H.E., Skrutskie M. 1998 PASP 110 367\\
Bohlin R.C., Savage B.D. and Drake J.F. 1978 ApJ 132\\
Boyle B. J., Shanks, T., Croom S.M., Smith R.J., Miller L., Loaring N., Heymans C. 2000 MNRAS 317 1014 \\
Braccesi A., Lynds R., Sandage A. 1968 ApJ 152 105\\
Chen H.-W. 2001 arXiv:astro-ph/0108171\\
Coleman G.D., Wu C.-C. and Weedman D.W. 1980 ApJS 43 393\\
Crampton D., Cowley, A.P., Hartwick F.D.A., Ko P.W. 1992 AJ104 1706\\
Crampton, Cowley, Schmidtke, Janson and Durrell 1988 AJ 96 816 \\
Croom S.M., Warren S.J., Glazebrook K. arXiv:astro-ph/0107451\\
Ellison S.L., Yan L., Hook I.M., Pettini M., Wall J.V., Shaver P. 2001 A\&A 379 393\\
Fall S.M., Pei Y.C. 1995 ApJ 454 69\\
Fall S.M., Pei Y.C., McMahon R.G. 1989 ApJ 341L 5\\
Firth A.E. 2001 arXiv:astro-ph/0108182\\
Francis, P.J., Whiting, M.T., and Webster, R.L. 2000, PASA, 53, 56 \\
Francis P.J. \etal, 1991, ApJ, 373, 465\\
Gunn J.E., Stryker L.L. 1983 ApJS 52 121\\
Hewett and Foltz 1994 PASP 106 113 \\
Hooper E.J., Impey C.D., Foltz C.B. and Hewett P.C. 1995 ApJ 445 62\\
Hooper E.J., Impey C.D. and Foltz C.B. 1997 ApJ 480 L95\\
Ives D.J., Tulloch S, Churchill J. 1996 SPIE 2654 266\\
Kochanek C.S. 1996 ApJ 466 638\\
Mackay, C.D., \etal 2000, SPIE, 4008, 1317\\
Madau P. 1995 ApJ, 441, 18 \\
Madgwick D.S. \etal 2001 arXiv:astro-ph/0107197\\
Magorrian \etal 1998 AJ, 115 2285\\
Masci F.J. Webster R.L., Francis P.J. 1998 MNRAS 301 975\\
McCarthy P., Chen H.-W., Martini P., Persson S.E., Oemler A., Carlberg R.,Abraham R.,Firth A., McMahon R., Lahav O., Sabbey C., Marzke R., Ellis R., Somerville R., Wilson J. 2001 AAS 198 7901 \\
McMahon R.G., WaltonN.A., Irwin M.J., Lewis J.R., Bunclark P.S., Jones D.H. 2001 NewAR 4597M\\
P\'{e}roux C., Storrie-Lombardi L.J., McMahon R.G., Irwin M.,Hook I.M. 2001 AJ 121 1799\\
Prevot \etal 1984 A\&A 132 389\\
RichardsG.T. \etal 2001AJ 122 1151\\
Sabbey, C.N., McMahon, R.G., Lewis, J.R., and Irwin, M.J. 2001, ASP Conference, ADASS X, (arXiv:astro-ph/0101181)\\
Seaton M.J. 1979, MNRAS, 187, 73\\
Sharp R.G., McMahon M.G., Irwin M.J. and Hodgkin S.T. 2001 MNRAS 326L 45\\
Storrie-Lombardi L.J., McMahon R.G., Irwin M.J.  1996 MNRAS 283L 79\\
Vanden Berk D.E., SDSS Collaboration arXiv:astro-ph/0105231\\
Veron-Cetty M.P., Veron P., 2000, ESO Scientific Report 19, 1 \\
Walton N.A., Lennon D.J., Irwin M.J., McMahon R.G. 2001 INGN 4 3\\
Warren, S.J., Hewett, P.C., and Foltz, C.B. 2000, MNRAS, 312, 827 \\
\end{document}